\documentclass{emulateapj}
\usepackage{graphicx}

\slugcomment{submitted to the Monthly Notices of the Royal Astronomical Society}

\shorttitle{Optical Emission from G182.4+4.3}
\shortauthors{Fesen et al.}

\begin{document}

\title{Detection of Extensive Optical Emission from the Extremely Radio Faint 
Galactic Supernova Remnant G182.4+4.3}

\author{ Robert A.\ Fesen\altaffilmark{1},
        Jack M.\ M.\ Neustadt\altaffilmark{1,2},
        .Thomas G.\ How\altaffilmark{3}, and 2
        Christine S.\ Black\altaffilmark{1}
 }

\altaffiltext{1}{Department of Physics \& Astronomy, Dartmouth
                 College, Hanover, NH 03755 USA}
\altaffiltext{2}{The Ohio State University, Columbus, Ohio 43210, USA}
\altaffiltext{3}{Curdridge Observatory, Southampton UK }

\begin{abstract}

Wide-field H$\alpha$ images of the radio faint Galactic supernova remnant
G182.4+4.2 reveal a surprisingly extensive and complex emission structure, with
an unusual series of broad and diffuse filaments along the remnant's
southwestern limb. Deep [\ion{O}{3}] 5007 \AA \ images reveal no appreciable
remnant emission with the exception of a single filament coincident with the
westernmost of the broad southwest filaments.  The near total absence of
[\ion{O}{3}] emission suggests the majority of the remnant's optical emission
arises from relatively slow shocks ($\leq$70 km s$^{-1}$), consistent with
little or no associated X-ray emission.  Low-dispersion optical spectra of
several regions in the remnant's main emission structure confirm a lack of
appreciable [\ion{O}{3}] emission and indicate [\ion{S}{2}]/H$\alpha$ line
ratios of $0.73 - 1.03$, consistent with a shock-heated origin. We find G182.4+4.2 to be a
relatively large (d $\sim$50 pc at 4 kpc) and much older (age $\sim$40 kyr)
supernova remnant than previously estimated, whose weak radio and X-ray
emissions are related to its age, low shock velocity, and location in a low
density region some 12 kpc out from the Galactic centre.  

\end{abstract}
\bigskip


\section{Introduction}

The majority of the  nearly 300 currently identified Galactic supernova
remnants (SNRs) were discovered at radio wavelengths due to their
characteristic nonthermal radio emission associated with shocked gas.  Of
these, only some 30\% exhibit any coincident optical emission.

Among the faintest radio Galactic SNRs known is G182.4+4.3.  It was discovered
by \citet{Kothes1998} using the Effelsberg 100 m radio telescope at frequencies
of 1400, 2675, 4850, and 10450 MHz who found it to have a spectral index
$\alpha = -0.42 \pm 0.10$ and a polarization percentage exceeding 60\% for the
brightest southwestern parts of its emission shell.  

Based on the remnant's bright southwestern limb at 1400
and 2675 MHz and a partially complete spherical shell weakly seen at 4850 MHz,
\citet{Kothes1998} estimated a diameter of $50^{\prime}$.  They also estimated
a distance of at least 3 kpc, an age of 3800 yr, and a shock velocity of 2300 km
s$^{-1}$ based upon an estimated relatively low ambient interstellar density, n$_{0}$, of
less than 0.02 cm$^{-3}$ and a Sedov expansion model with a ratio of 5 to 10
for swept-up mass to ejected mass.

Follow-up observations made by \citet{Reich2002} at 2.7 and 4.9 GHz and
by \citet{Sun2011} at 6 cm determined $\alpha = -0.41 \pm 0.14$.  CO observations
made by \citet{Jeong2012} showed the boundary of a molecular cloud off to the
remnant's northeast limb that matched its radio boundary although no
evidence for interaction was found.

Optical emission associated with the G182.4+4.3 was first reported by
\citet{Sezer2012}. Four $13.5^{\prime} \times 13.5^{\prime}$ regions were imaged
using on and off H$\alpha$ and [\ion{S}{2}] 6716,6731 narrow passband filters.
These images, which covered portions of the remnant's centre, southern, northern,
and northwestern regions revealed both filamentary and diffuse emission which
they found was correlated with the remnant's radio structure.  

The optical filaments exhibited [\ion{S}{2}]/H$\alpha$ ratios of 0.9 to 1.1 $\pm
0.1$ based on flux-calibrated images, well above the standard SNR
[\ion{S}{2}]/H$\alpha$ ratio criteria of 0.4 
\citep{MC72,Dodorico80,Blair81,Dopita84}, confirming the 
shocked nature of the detected optical emission.  \citet{Sezer2012} also
reported finding associated X-ray emission based on {\sl XMM-Newton} data and
estimated a relatively young age of just 4400 yr assuming a distance of 3
kpc.

No associated optical emission is visible on the emission-line survey of the
Galactic plane of \citet{Parker79} and the remnant's Galactic region was not
surveyed by the much deeper Virginia Tech Spectral Line Survey (VTSS) of the
Galactic Plane \citep{Dennison98,Fink03}. However, while a few  of the
remnant's brighter filaments are visible in the Isaac Newton Telescope (INT)
Photometric H$\alpha$ Survey (IPHAS) of the Northern Galactic Plane
\citep{Drew05,GS08}, much of G182.4+4.3's overall optical structure is surprisingly
visible on the digitized broadband red image of the
second Palomar Sky Survey (DSS2) (see Fig.\ 1).

Here we report a deep mapping of G182.4+4.3's  optical emission structure.
Wide-field H$\alpha$ images reveal an extensive optical structure from this
extremely radio faint SNR, including considerable interior emission and 
a remarkable set of broad and diffuse optical shock filaments along its
southwestern limb, plus a near total absence of [\ion{O}{3}] line emission.  

Our optical imaging data and follow-up low-dispersion
spectra are described in $\S$2, with results presented in $\S$3.
Combining these new data with prior optical and radio observations, we discuss
in $\S$4 the remnant's general properties including its distance, size, and
shock velocity.  Our conclusions are summarized in $\S$5.

\begin{figure*}[t]
\includegraphics[angle=0,width=17.0cm]{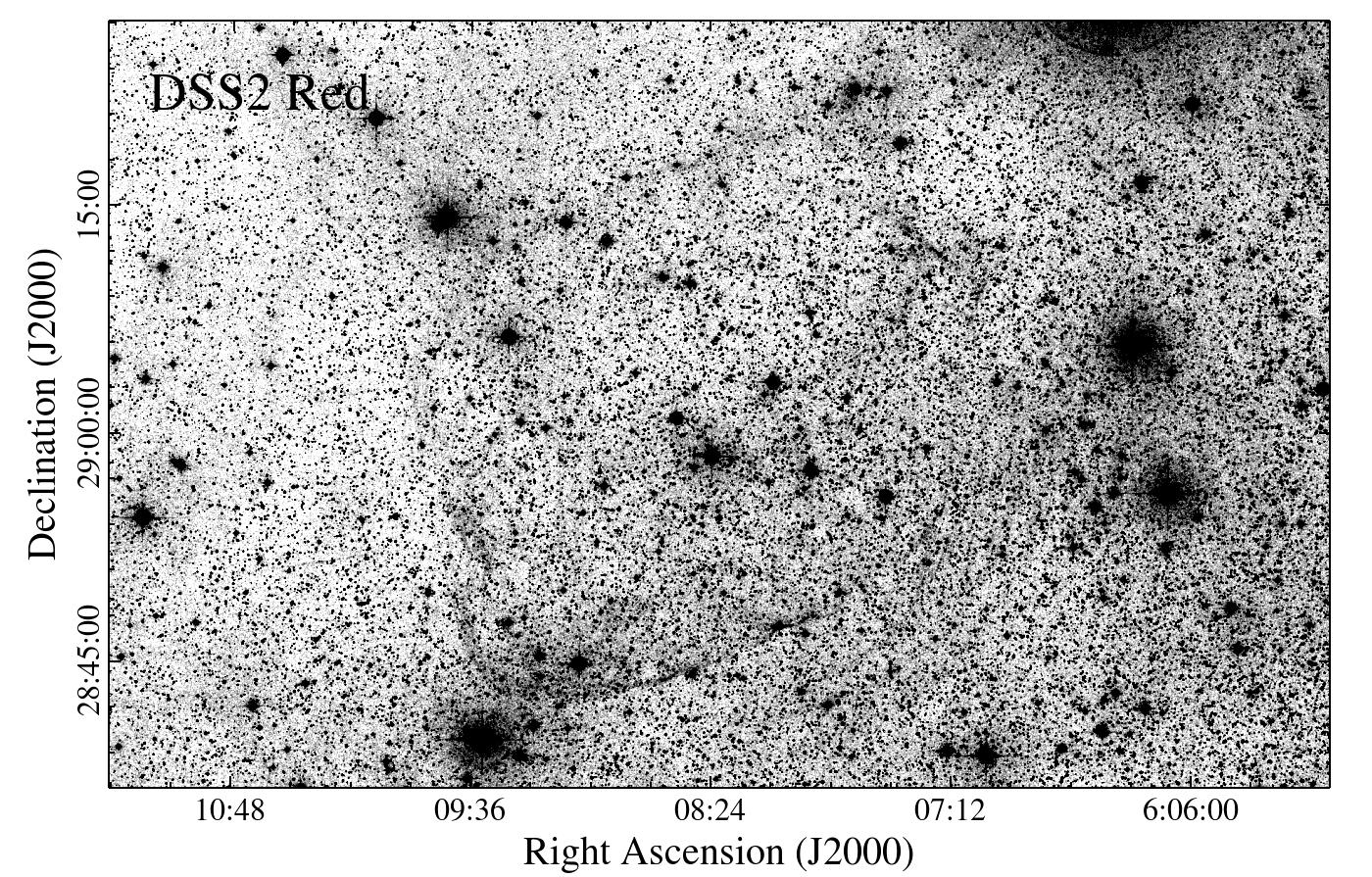}
\includegraphics[angle=0,width=17.0cm]{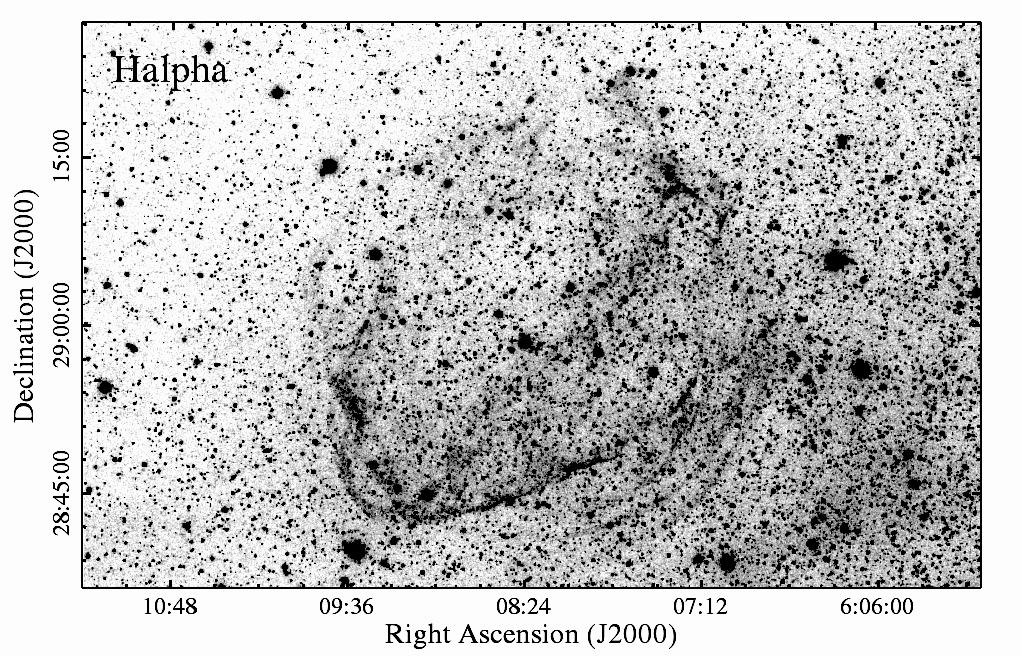}
\caption{{\bf{Upper Panel:}} A smoothed reproduction of the digitized red Palomar Sky Survey image (DSS2) 
revealing emission associated with the SNR G182.4+4.3. 
(A faint satellite trail running across the remnant has been
digitally removed.)
{\bf{Lower Panel:}} A deep H$\alpha$ image of the SNR G182.4+4.3 showing a more extensive
optical emission structure along with an unusual 
series of broad filaments along its southwestern limb which is only hinted at in the DSS2 image. 
Note that the DSS2 image shows a more complete line of emission along the remnant's 
northern boundary than seen in the H$\alpha$ image. }
\label{How_image}
\end{figure*}

\begin{figure*}[t]
\begin{center}
 \includegraphics[width=17.0cm]{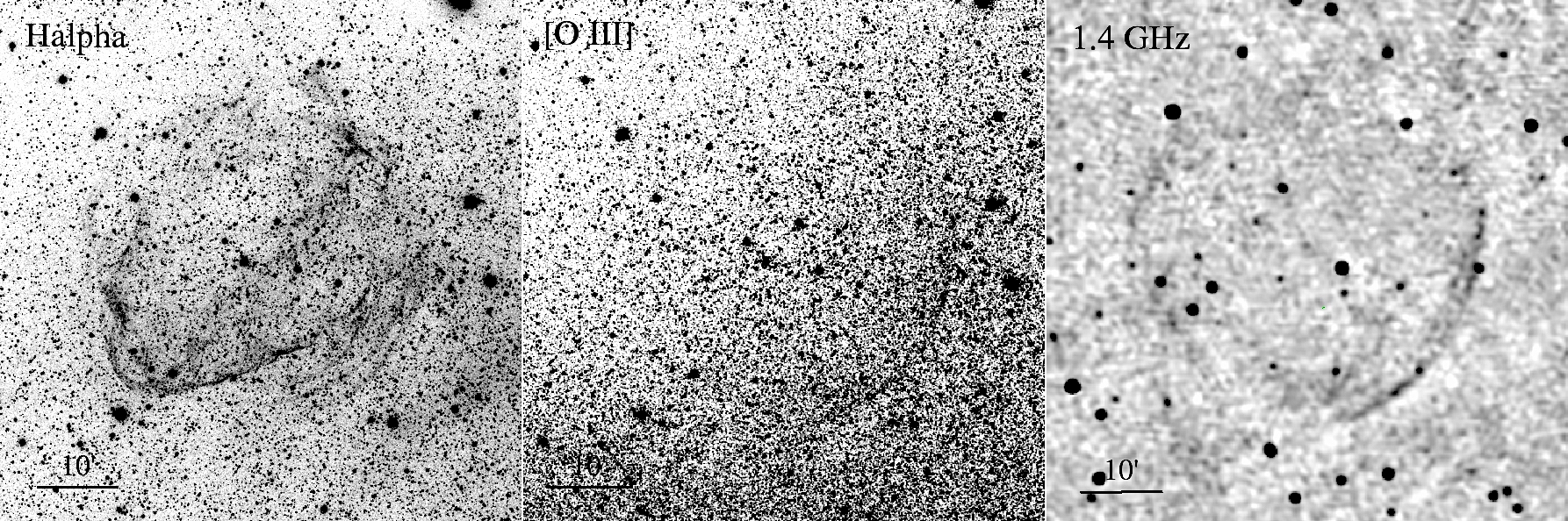} \\
 \includegraphics[width=17.0cm]{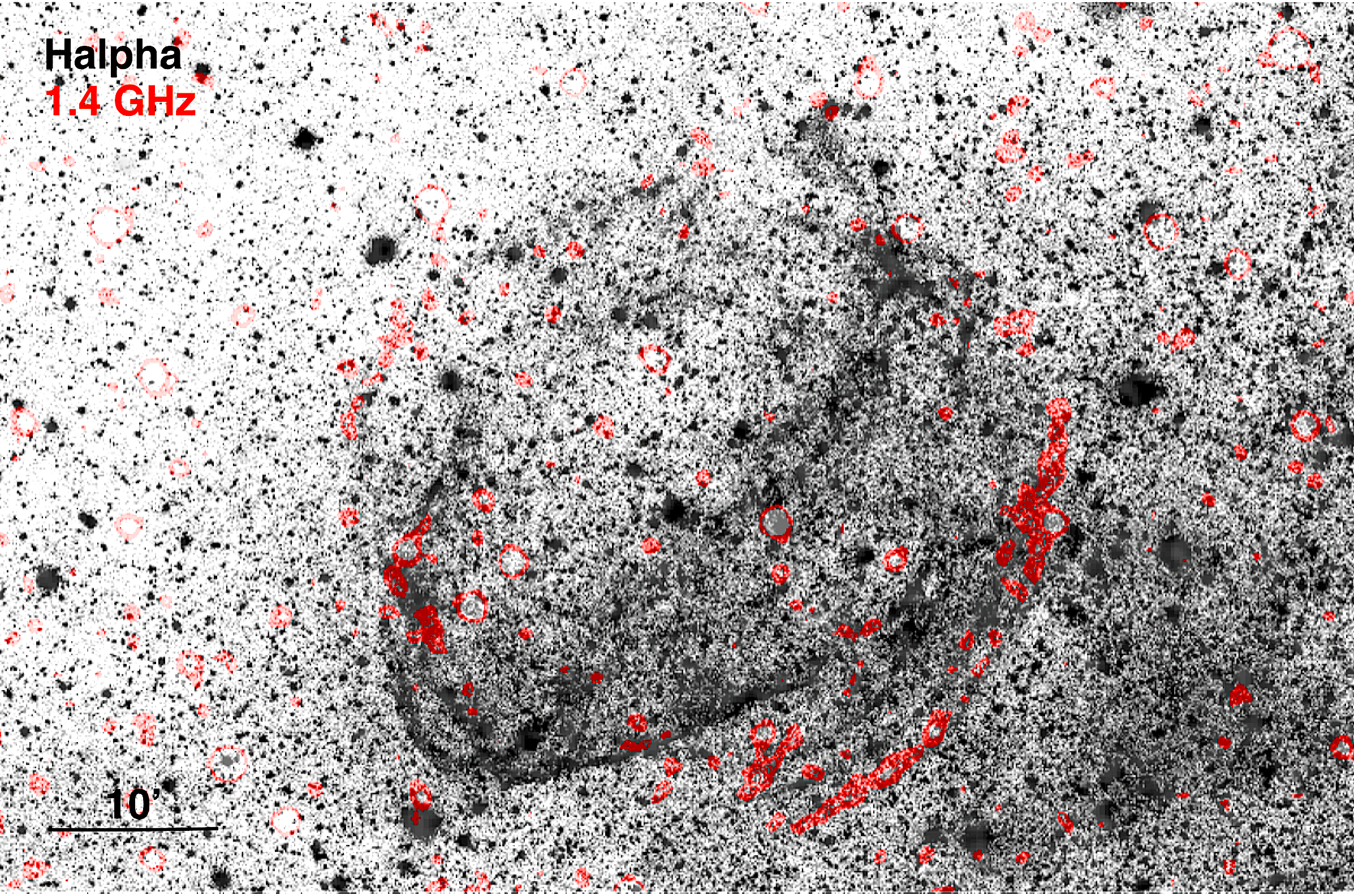}
\caption{  {\bf{Upper Panels:}} Comparison of emissions from G182.4+4.3: H$\alpha$ (left), 
[\ion{O}{3}] (middle), and 1.4 GHz radio (right).
While the remnant exhibits extensive emission in H$\alpha$, it is virtually 
undetectable in [\ion{O}{3}] 5007 \AA \ except for a single
faint filament located near the outermost of the 
H$\alpha$ filaments in the southwest. This [\ion{O}{3}] filament
is coincident with the brightest remnant feature seen in the radio.
  {\bf{Lower Panel:}} Overlay of the 1.4 GHz radio image (red) onto the Halpha image (black). 
 }
\end{center}
\label{three_images}
\end{figure*}


\begin{figure*}[t]
\begin{center}
\leavevmode
\includegraphics[width=17.0cm]{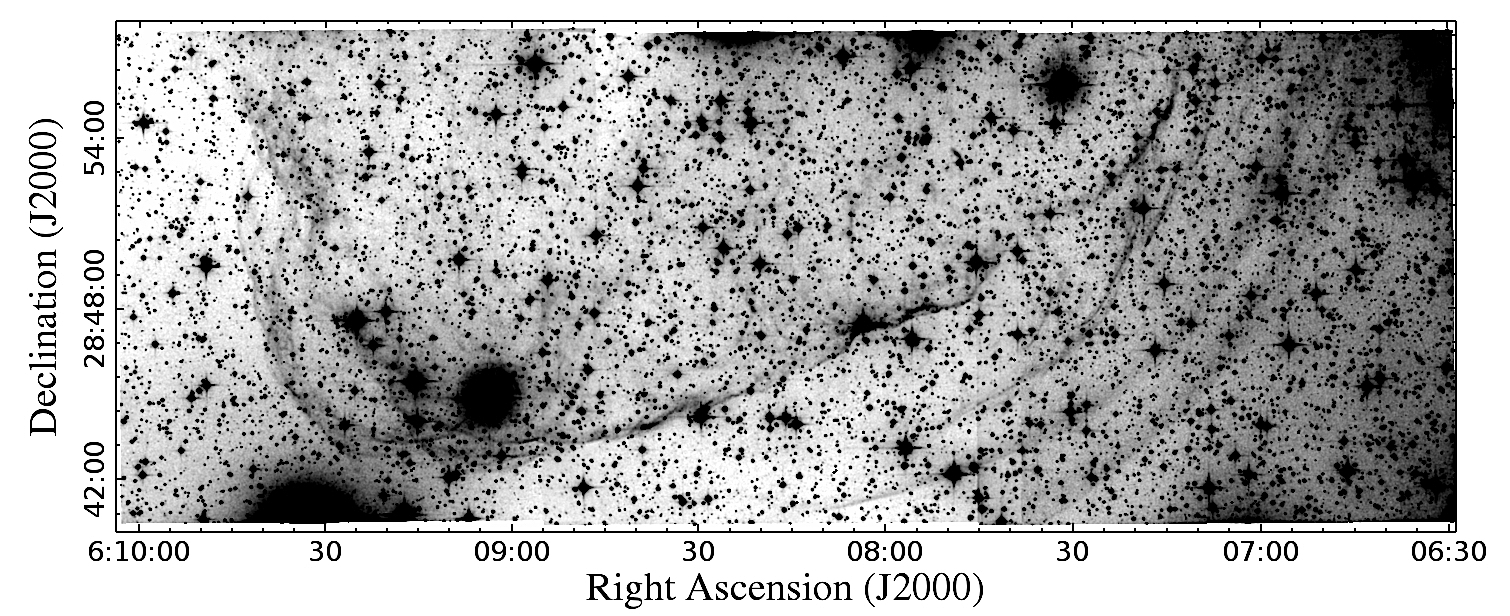}
\hspace{1em}
\includegraphics[width=17.0cm]{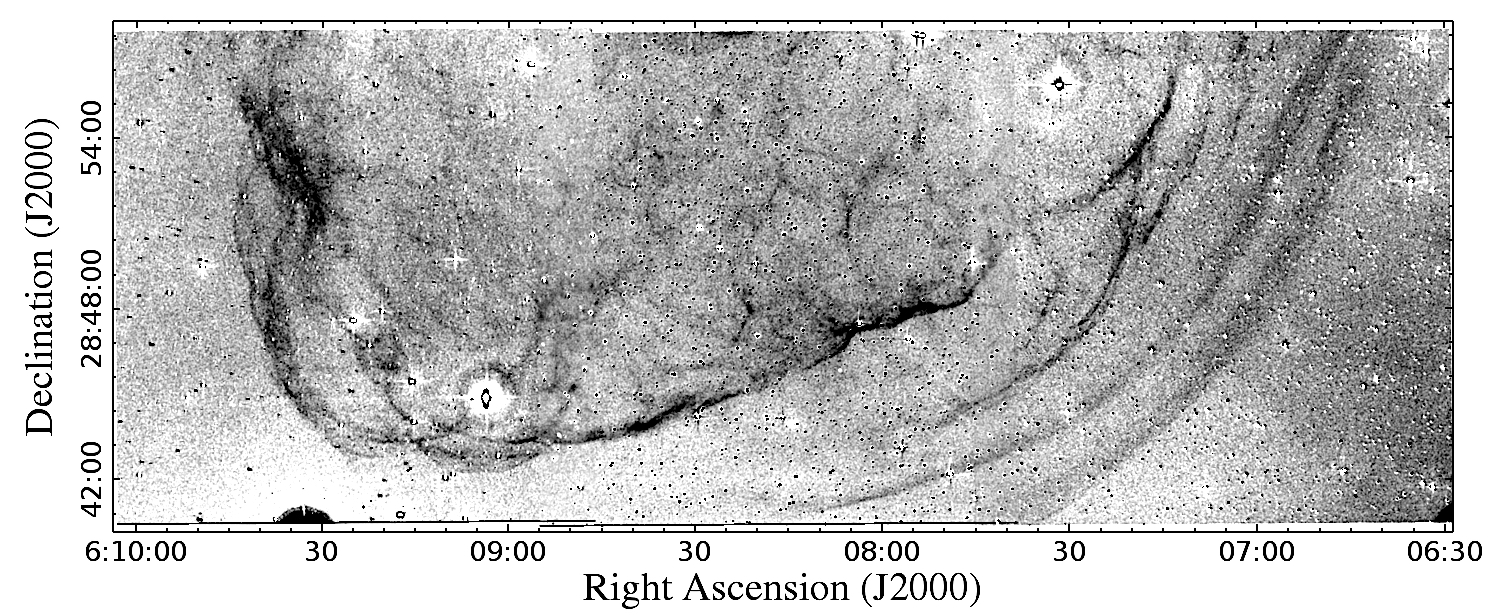}
\caption{Mosaics of H$\alpha$ images (upper panel) and H$\alpha$ continuum 
subtracted images (lower panel)  of G182.4+4.3's
southern limb region. Note the bright filaments marking the remnant's 
southern boundary, the numerous short curved filaments
in the remnant's interior, and the set of unusually broad diffuse  
filaments situated farther to the southwest.  
}
\end{center}
\label{South_Region}
\end{figure*}

\begin{figure*}[t]
\begin{center}
\leavevmode
\includegraphics[width=17.0cm]{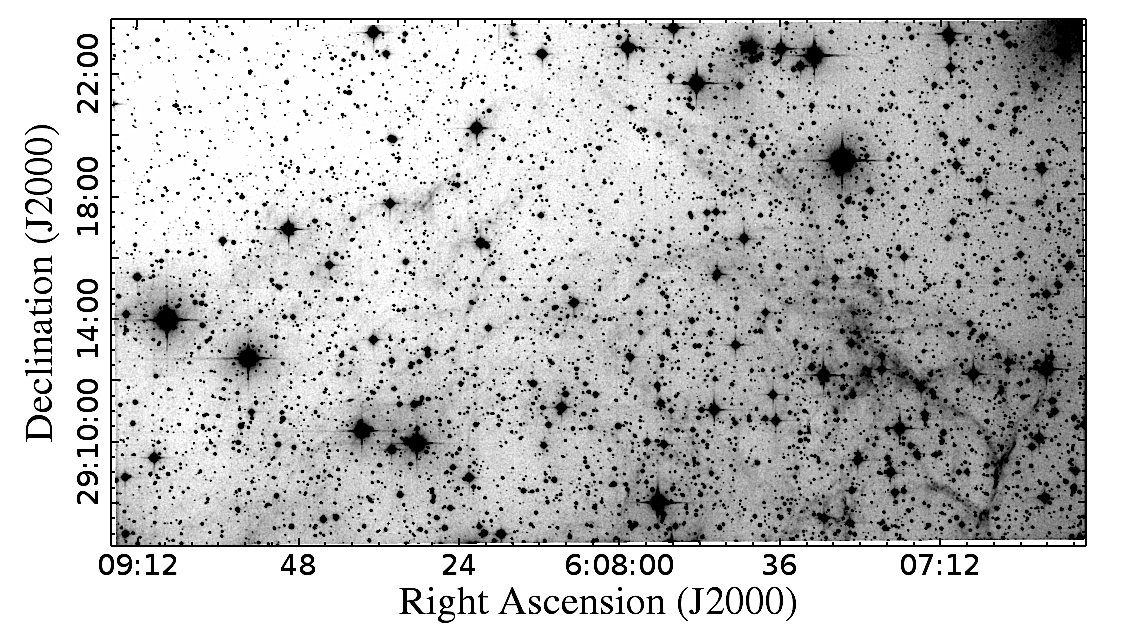}
\hspace{1em}
\includegraphics[width=17.0cm]{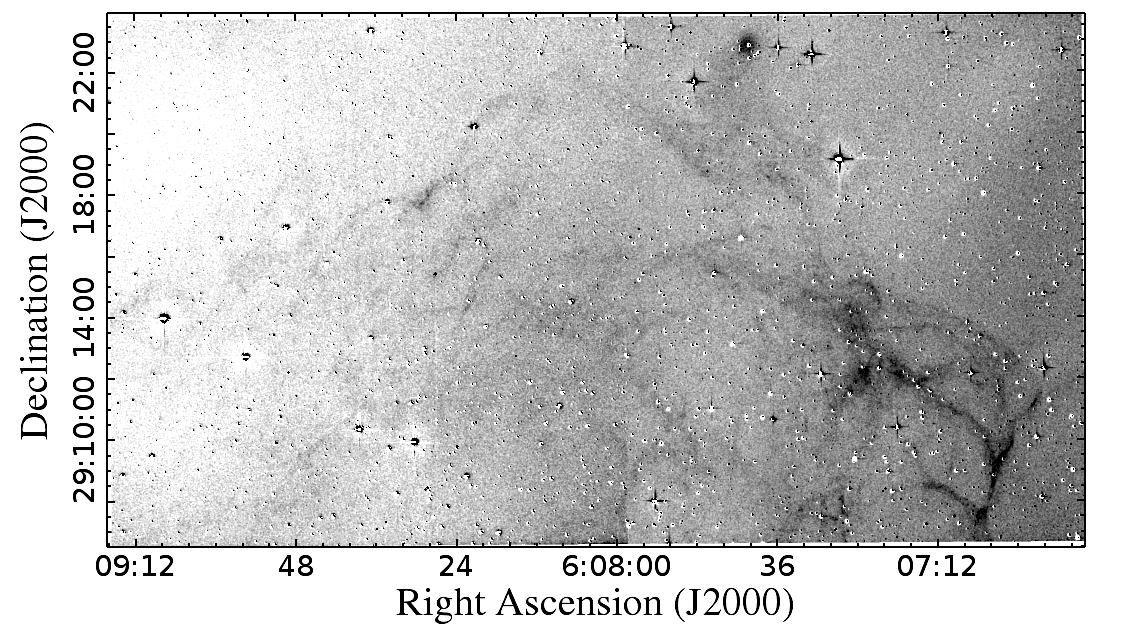}
\caption{Mosaics of H$\alpha$ images (upper panel) and H$\alpha$ 
continuum subtracted images (lower panel) of G182.4+4.3's  
northern limb region. 
  }
\end{center}
\label{North_Region}
\end{figure*}

\begin{figure*}[h]
 \begin{center}
    \leavevmode
    \includegraphics[scale=0.49]{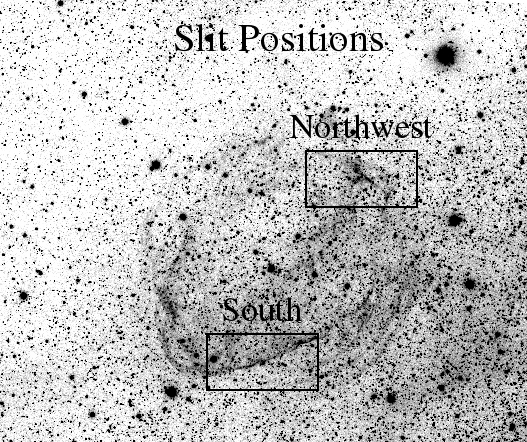}
    \includegraphics[scale=0.49]{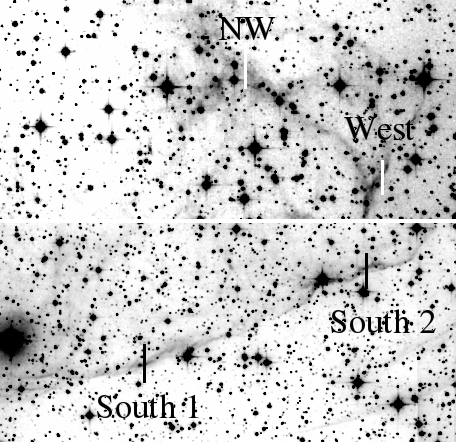}
    \includegraphics[width=0.99\linewidth]{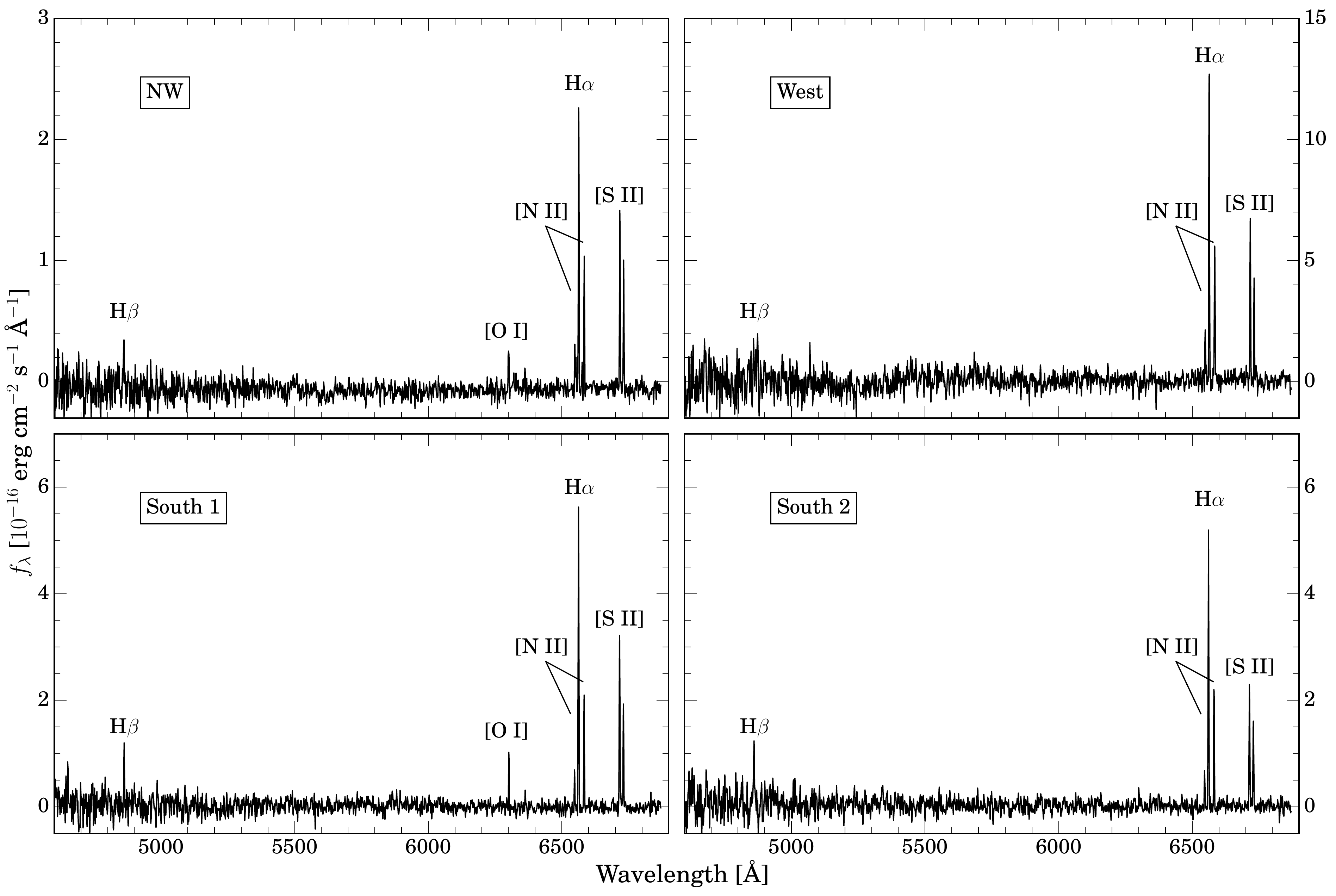}
\caption{ {\bf{Upper panels:}} Locations of the four slit positions.  {\bf{Lower panels:}} Optical spectra for the four slit positions. }
\end{center}
\label{Slits_n_spectra}
\end{figure*}

\section{Observations}

Wide-field images centred on the remnant G182.4+4.3 were obtained in February
2016 with a f/4 90 mm refractor at the Curdridge Observatory in Southampton,
UK.  The images were taken using an Astrodon 30 \AA \ FWHM  H$\alpha$ filter and an
Atik 490 camera employing a Sony ICX814 CCD which was binned 2 $\times$ 2
yielding $1690 \times 1352$ pixels.  This system provided a FOV of $64^{\prime}
\times 86^{\prime}$ with an image scale of $3.72^{\prime\prime}$ pixel$^{-1}$.

By combining 25 dithered 1500 s exposures for a total exposure time of 10.4 hr,
faint but extensive H$\alpha$ emission was detected within the boundaries of
the remnant's radio emission.  Similarly long 30 \AA \ FWHM [\ion{O}{3}] filter
exposures showed little coincident emission with the exception of a single 
filament along its southwestern limb.

Based on these images, low-dispersion spectra of four filaments in the
remnant's northern, northwestern and southern regions were subsequently
obtained with the 2.4m Hiltner telescope at MDM Observatory in January and
Feburary 2017 using the OSMOS Spectrograph \citep{Martini2011}. Using a blue
VPH grism (R = 1600) and a 1.2 arcsec N-S slit, exposures of $2 \times 1200$ s
and $2 \times 2000$ s were taken of four filament regions covering 3900--6900
\AA \  with a spectral resolution of 1.2 \AA \ pixel$^{-1}$. Spectra were
extracted from regions between 4 and 9 arcseconds along each of the slits.

Follow-up, higher resolution direct images were subsequently obtained on 22
February 2017 using the 1.3 m McGraw-Hill telescope at the MDM Observatory at
Kitt Peak outside of Tucson, Arizona.  A series of narrow passband H$\alpha$
and red continuum 6450 \AA \ images (FWHM = 90 \AA) with exposure times of 900
s were taken covering most of the remnant's northern and southern regions using
a 2k $\times$ 2k SITe CCD providing a FOV of $17^{\prime} \times 17^{\prime}$,
with on-chip binning yielding an image scale of $1.06^{\prime\prime}$
pixel$^{-1}$.

Standard pipeline data reduction of MDM images and spectra using
IRAF\footnote{IRAF is distributed by the National Optical Astronomy
Observatories, which is operated by the Association of Universities for
Research in Astronomy, Inc.\ (AURA) under cooperative agreement with the
National Science Foundation.} was preformed.  Spectra were similarly reduced
using IRAF and the software L.A. Cosmic \citep{vanDokkum2001} to remove cosmic
rays.  Spectra were calibrated using an Ar lamp and spectroscopic standard
stars \citep{Oke74,Massey90}.

\section{Results}

\subsection{Images} 

G182.4+4.3's full H$\alpha$ emission structure can been seen in the wide-field,
low resolution images shown in Figure~1.  Despite its extremely faint radio
emission, the remnant exhibits an extensive optical emission structure,
brightest in the south, southeast and northwest, with considerable and
unusually bright interior emission.  A circle 43$^{\prime}$ in diameter centred
at $\alpha$(J2000) = 06:08:22.5, $\delta$(J2000) = +28:58:45 ($l$ =
182.46$^{\rm o}$, $b$ = $+4.31^{\rm o}$) can encompass all of the remnant's
optical emission except for a small extension off the remnant's northwestern
limb. 
 
In contrast to its extensive H$\alpha$ emission, our deep [\ion{O}{3}] 5007 \AA
\ image shows G182.4+4.3 to emit almost no [\ion{O}{3}] line emission. 
Comparison of the remnant's H$\alpha$ and [\ion{O}{3}] images, along with a 1.4
GHz NRAO VLA Sky Survey (NVSS; \citealt{Condon1998}) radio emission map is
presented in Figure 2. 

As can be seen in this figure, while the remnant shows considerable H$\alpha$
emission, our [\ion{O}{3}] image detected only a single [\ion{O}{3}]
emission filament.  Although having nearly the same curvature as this
westernmost H$\alpha$ filament, the [\ion{O}{3}] filament is not precisely
coincident with it, but is instead offset farther to the west some
$20^{\prime\prime} - 30^{\prime\prime}$ from the H$\alpha$ filament. It also
appears to extends a bit farther northward than that of the H$\alpha$ filament.

\cite{Sezer2012} claimed the presence of {\sl XMM-Newton} detected X-ray
emission from an extended region some 16 arcminutes in diameter located along
the the remnant's southwestern remnant centred on RA=06:07:10, Dec=+28:52:05
near and behind (East) of the location of the remnant's sole [O III] bright
filament.  However, examination of {\sl ROSAT} All-SKY Survey (RASS3; 0.5-2.0
keV) data could not confirm the presence of any appreciable X-ray
associated emission with the remnant. 

The lower panel of Figure 2 reveals a close correlation of the remnant's
brightest radio emission filament with its outermost southwestern H$\alpha$
filament, along with a faint H$\alpha$ filament in its eastern limb perhaps
best seen in the DSS2 image shown in Figure 1, and the relatively bright
complex of overlapping H$\alpha$ filaments also in the East.  We note that
[\ion{O}{3}] filament also roughly matches the position and extent the
remnant's bright southwestern 1.4 GHz radio emission filament.

Higher resolution views of the remnant's southern and northern limbs are
presented in the image mosaics of Figures 3 and 4. H$\alpha$ and continuum
subtracted H$\alpha$ images are shown for both regions.  Although
\citet{Sezer2012} also imaged part of the remnant's northwestern and southern
limbs, more emission can be seen in these figures, especially in
regard to interior emission. From their image data, \citet{Sezer2012} found
diffuse interior emission especially in the north and central regions. However,
our continuum subtracted images instead reveal a complex structure of short, faint
and curved interior filaments that could be mistaken as diffuse
emission.

While much of G182.4+4.3's H$\alpha$ emission structure is not exceptional
compared to that seen in many other Galactic SNRs, the presence of so much
optical emission for such a radio faint SNR is wholly unexpected. Moreover, the
morphology of its broad and diffuse H$\alpha$ filaments located further south
and west of the remnant's nominal emission structure is highly unusual.
Whereas a continuous line of relatively bright filaments stretch from the
remnant's southeastern corner to its western limb that would seem to mark its
southern boundary, several long and parallel filaments are seen considerably
farther west.
  
These outer southwestern filaments give a sense of concentric shock fronts
outside of the remnant's otherwise seemingly well defined southwest boundary.
They are also unusually broad and partially diffuse in appearance, exhibiting a
brightness gradient  with increasing distance behind their leading (western)
edges. As noted above, the western edge of the more western filament is close
to but not exactly coincident with the remnant's sole [\ion{O}{3}] filament.

The remnant's northern H$\alpha$ emission (Fig.\ 4) is generally fainter than
in the south (Fig.\ 3), with the exception of a complex of brighter filaments
along its northwestern limb. These filaments were partially imaged by
\citet{Sezer2012}.  As can be seen in the continuum subtracted image and in the
red DDS2 image (Fig.\ 1), the remnant's H$\alpha$ emission is nearly continuous
along the northern limb.

\subsection{Optical Spectra}

Due to the relative faintness of the remnant's optical emissions,
low-dispersion spectra were only obtained at four regions on the remnant's
brighter northwestern and southern filaments.  Exact slit positions are
shown in the upper panels of Figure~5, with the resulting reduced spectra
shown in the figure's lower panels.  Spectra of these regions showed
[\ion{S}{2}]/H$\alpha$ ratios between 0.75 and 1.03 ($\pm 0.1$), consistent
with the earlier findings of \citet{Sezer2012} and well above the standard SNR
identification criteria of $\geq$0.4 \citep{Dodorico80,Blair81,Dopita84}.
 
In the four filament spectra, the ratio of the density sensitive [\ion{S}{2}]
6716, 6731 \AA \ emission lines are all near the low density limit of 1.43
(within measurement error), indicating postshock election densities $\leq$ 100
cm$^{-3}$ \citep{Oster06}. Also, consistent with the results of the direct imaging,
none of the four spectra showed any appreciable [\ion{O}{3}] 4959, 5007 \AA \ emission.  

Because [\ion{O}{1}] 6300, 6364 \AA \ emission is often seen in SNR
spectra \citep{Fesen1985}, the lack of [\ion{O}{1}] in the spectra taken at slit
positions South 2 and West deserves special comment.  Although imperfect
subtraction of the bright night sky [\ion{O}{1}] emission relative to the
faintness of the remnant's filaments could in principle account for the
apparent lack of [\ion{O}{1}] emission at these two slit locations, no such
issue was present for slit positions South 1 and Northwest (NW) where
[\ion{O}{1}] emission was clearly detected.  In addition, a careful examination
did not indicate background [\ion{O}{1}] sky emission subtraction problems at
either slit position.  Rather, background subtraction of extended [\ion{O}{1}]
emission from surrounding remnant filaments may have lead to the seeming
absence of [\ion{O}{1}] in these faint filaments. We note that the presence of
some filaments elsewhere in the remnant exhibiting relatively strong
[\ion{O}{1}] emission might account for some differences in the remnant's
morphology seen in the broadband red DSS2 image versus our H$\alpha$ image (see
Fig.\ 1). Hence, we conclude it likely that many, if not most, of the remnant's
filaments exhibit some [\ion{O}{1}] emission as expected from shocked
interstellar gas.  

We detected H$\beta$ emission in three of the filaments studied and thus can
estimate the reddening in these directions. The observed H$\alpha$/H$\beta$
ratios for the filamentary emissions at slit positions South 1, South 2, and NW
are 4.54, 3.40, and 4.35, respectively. Assuming a temperature of 10,000 K and
an intrinsic H$\alpha$/H$\beta$ ratio of 2.87, these values suggest a range of
$E(B-V)$ values between 0.17 and 0.43 (see Table 1). 

\section{Discussion}

G182.4+4.3's extensive and relatively bright optical structure is surprising 
given its very faint radio emission which places it in the bottom 1\% of the
almost 300 known Galactic SNRs \citep{Green2015}.  No other Galactic SNR with
such a low surface brightness near G182.4+4.3's value of $\Sigma_{1~ \rm GHz}$
= $7.5 \times 10^{-23}$ Watt m$^{-2}$ Hz$^{-1}$ sr$^{-1}$ \citep{Kothes1998}
shows nearly as extensive optical emission, if they show any detectable optical
emission at all \citep{Green2014}.

The remnant is also unusual in showing long, smoothly curved and parallel H$\alpha$
filaments along its southwestern and western limbs which are quite unlike that seen in any other
Galactic SNR. We speculate that these curiously parallel filaments may indicate a large-scale
blowout of the remnant's shock front in this direction.  Moreover, the presence of 
separate filaments may also indicate small ISM density variations along our line
of sight. 

The remnant's overall near total lack of appreciable [\ion{O}{3}] emission is
also relatively rare.  Optical spectra of Galactic and extragalactic SNRs
reveals only a handful of remnants that globally exhibit no 
[\ion{O}{3}] emission. In those cases, the
remnants are both physically large and estimated to be relatively old
\citep{Fesen1985,Long2018}.  

It is interesting to note that the remnant's sole [\ion{O}{3}] emission filament is not
exactly coincident with the outermost southwest H$\alpha$ filament. Instead, it
is displaced by some 20 to 30 arc seconds farther west yet in alignment in size
and extent with the remnant's brightest radio emission. This could be a sign of
significant post-shock cooling, signaling the remnant may be evolving out of
the adiabatic and into the radiative phase.

The presence of strong [\ion{S}{2}] and [\ion{N}{2}] emissions but weak or absent [\ion{O}{3}]
emission is a signature of relatively low velocity shocks ($\leq$70 km s$^{-1}$)
\citep{Raymond79,Shull79,Hart87}. However, such a conclusion is
contrary to earlier estimates for G182.4+4.3's properties where it has been
suggested to be a relatively young SNR with a shock velocity around 2000 km
s$^{-1}$ \citep{Kothes1998,Sezer2012}.  

Knowledge of G182.4+4.3 true physical size would greatly aid in determining its 
evolutionary phase but this is unknown due to its uncertain distance.
\citet{Kothes1998} estimated a distance of at least 3 kpc, whereas \citet{Gus03}
cites a value of 4.83 kpc and \citet{Pavlovic2013,Pavlovic2014} a value of 4.2
-- 4.4 kpc based on various versions of a $\Sigma-D$ relation. 

With an angular diameter of appromiately 43$^{\prime}$ based on our optical images,
G182.4+4.3 would have a linear diameter of around 38 pc if placed at its initial
estimated minimum distance of 3 kpc \citep{Kothes1998,Sezer2012}.  However, a
distance in excess of 3 kpc seems likely. 

One can use our derived $E(B-V)$ reddening values from our optical spectra to
help determine its distance through reddening values as a function of distance
from the three-dimensional maps presented in \citet{GGreen2015} which utilized
Pan-STARRS \citep{Schlafly2014} and 2MASS \citep{Skrutskie2006}
photometry\footnote{http://argonaut.skymaps.info}.  Spectra of remnant's
southern filaments showed $E(B-V)$ values between 0.17 and 0.43 (see Table 1).
This range is consistent with distances anywhere from 0.5 to 4.0 kpc. However,
given the weakness of and low S/N for H$\beta$ in our three spectra, the fact
that two filaments (NW and South 1) indicate similar $E(B-V)$ values of around
0.40 suggests a distance of 4 kpc or more.

If one takes the remnant to be in the adiabatic Sedov-Taylor phase of its
evolution, then its radius, R, in pc is equal to: 
12.5 E$_{51}^{1/5}$ n$_{\rm
o}^{-1/5}$ t$_{\rm 10000 yr}^{2/5}$ and its shock velocity, V$_{\rm shock}$, is
equal to: 1950 km s$^{-1}$ E$_{\rm 51}^{1/5}$ n$_{\rm o}^{-1/5}$ t$_{\rm 1000
yr}^{-3/5}$ where  E$_{\rm 51}$ is the SN energy release in units of 10$^{51}$ erg
and n$_{\rm o}$ is electron density in cm$^{-3}$  \citep{Blondin1998}.  Adopting a distance of 3 kpc
and hence a linear radius of 19 pc, E$_{\rm 51}$ = 1, and n$_{\rm o}$ = 0.02
cm$^{-3}$ as estimated from radio observations \citep{Kothes1998,Sezer2012},
one finds an age $\sim$4000 yr and a shock velocity $\sim$1850 km s$^{-1}$.
Similar age and shock velocities for G182.4+4.3 were estimated by
\citet{Kothes1998} and \citet{Sezer2012}.  

However, such a young age and relatively high shock velocity appear
inconsistent with the remnant's overall emission properties.  Namely, its near
total absence of optical [\ion{O}{3}] line emission which points to a
relatively slow, not fast, shock velocity and its weak radio and X-ray fluxes. 

Moreover, if the remnant had a shock velocity of $\sim$2000 km s$^{-1}$, one would
also expect to find widespread nonradiative Balmer dominated filament emissions
like those seen in the remnants of Tycho's SNR (SN 1572) and SN 1006
\citep{Kirshner1987,Ghava2002} especially along it limbs. Instead we find, as
did \cite{Sezer2012}, ordinary shocked SNR optical emission filaments with
strong emission from low ionization states such as [\ion{N}{2}] and [\ion{S}{2}].

On the other hand, if the remnant were at a distance around 4 kpc, as suggested based
on $\Sigma - D$ estimates, the remnant would be roughly some 50 pc in diameter and
hence a larger, older and more evolved SNR. This would also put it more in
line with its lack of appreciable [\ion{O}{3}] emission due to expected lower filament
shock velocities.

If one adopts an ISM density of 0.5  cm$^{-3}$ and assumes the remnant is in
the Sedov phase, then a diameter of 50 pc and a SN energy of $1 \times
10^{51}$ erg leads to an estimated age of 40,000 yr and a blast wave velocity
$\sim$250 km s$^{-1}$.  This would mean the remnant is older and more
slowly expanding than previously estimated.

The remnant's overall optical line emissions also suggest a distance in excess
of 3 kpc and hence a fairly old and large remnant.  Compared to observed optical line
ratios seen in other Galactic SNRs \citep{Gon1983,Fesen1985}, our measured G182.4+4.3
values for [\ion{N}{2}]/H$\alpha$ ratio of between 0.5 and 0.6 and 
[\ion{S}{2}] 6716/6731 implying a low interstellar density (see Table 1)
are consistent with ai relatively large remnant (d $\sim$50 pc) and a
galactocentric distance of between 11 and 14 kpc.
The combination of a [\ion{N}{2}]/H$\alpha$ ratio around 0.55 and [\ion{S}{2}] 6716/6731 ratios
at the low density limit of 1.43 place G182.4+4.3 at the extreme range of such values for Galactic
SNRs (see Fig.\ 6 in \citealt{Fesen1985}).

Such an old, large remnant with a comparatively low shock velocity throughout much of
its structure is also more in line with our observations regarding [\ion{O}{3}] emission. 
Given its lack of [\ion{O}{3}] emission, a shock velocity of $50 -
60$ km s$^{-1}$ could be responsible for the observed optical spectrum seen for
the majority of its optical filaments. 

The exception is the remnant's [\ion{O}{3}] bright filament along its
southwestern limb, coincident with the remnant's brighest radio emission.
Without having a spectrum of it we cannot estimate its shock velocity. However,
models indicate that shocks above 80 km s$^{-1}$ are required to generate
appreciable [\ion{O}{3}] line emission \citep{Raymond79,Shull79,Hart87}.  The
cause for a higher shock velocity along the remnant's southwestern limb might
be a lower density region that the remnant is expanding into.

In old SNRs, optical filaments typically have a preshock density
several times the average ambient density, e.g., $1 - 10$ cm$^{-3}$.  Equating
ram pressures where $\rho_{1}$ $v_{1}^{2}$ = $\rho_{2}$ $v_{2}^{2}$, a 250 km
s$^{-1}$ shock moving through a 0.5  cm$^{-3}$ medium will drive a 65 km
s$^{-1}$ shock through a cloud of density $\sim$7 cm$^{-3}$.  

We note that with a size and age around 50 pc and 40 kyr, respectively,
G182.4+4.3 might have evolved out of its adiabatic phase and into the
radiative or snowplow phase. This could help explain its relatively extensive 
optical emission despite its weak radio flux \citep{Cox1972}.  In any case, we
suggest that G182.4+4.3 is not a young SNR but instead a relatively
large, old and well evolved SNR with weak emissions due to its old age, relatively
low shock velocity, and location in a relatively low density interstellar
medium some $300$ pc above the Galactic plane and 12 kpc out from the Galactic
centre.

\begin{deluxetable}{lrrrr}
\tablecolumns{5}
\tablecaption{Relative Fluxes in G182.4+4.3}
\tablewidth{0.95\linewidth}
\tablehead{
\colhead{Value} & \colhead{South 1} &  \colhead{South 2} & \colhead{NW} & \colhead{West} }
\startdata
~ RA (J2000)         & 6:08:37.90  & 6:07:55.31  & 6:07:18.53  & 6:07:01.71 \\
~ Dec(J2000)         & +28:43:52    & +28:47:57    & +29:12:27    & +29:09:37   \\
~ H$\beta$ 4861      &  22          & 29           &  23          &  $<10$ \\
~ [O III] 5007      &  $<10$       & $<15$        & $<15$        &  $<15$  \\
~ [O I] 6300         & 19           &  $<10$       & 12           &  $<15$       \\
~ [N II] 6548        & 13           & 12           & 14           & 16   \\
~ H$\alpha$ 6563     & 100          & 100          & 100          & 100   \\
~ [N II] 6583        & 39           & 39           & 45           & 45  \\
~ [S II] 6716        & 51           & 42           & 63           & 50   \\
~ [S II] 6731        & 30           & 31           & 43           & 32   \\
~ [N II]/H$\alpha$   & 0.52         & 0.51         & 0.59         & 0.61  \\
~ [S II]/H$\alpha$   & 0.81         & 0.73         & 1.06         & 0.82  \\
~ 6716/6731   & 1.7          & 1.35         & 1.47         & 1.6     \\
~ $\rho$ (cm$^{-3}$) & $\leq$100  & $\leq$100  &  $\leq$100  &  $\leq$100    \\
~ $E(B-V)$           & 0.43       & 0.17       & 0.39           & \nodata     \\
~ H$\alpha$ flux\tablenotemark{a} & 17.6  & 16.9   & 7.4   & 39.6    \\
\enddata
\tablenotetext{a}{Flux units: $10^{-16}$ erg s$^{-1}$ cm$^{-2}$. }
\end{deluxetable}

\section{Conclusions}

We report the presence of a surprisingly extensive optical emission structure associated
with the extremely faint radio supernova remnant G182.4+4.2. Deep H$\alpha$ images
reveal a complex and nearly complete emission shell $\approx$43$^{\prime}$ in
diameter, a set of unusually broad and diffuse filaments outside the remnant's
southwestern limb, and an almost complete lack of measurable [\ion{O}{3}]
emission with the exception of a single long filament coincident with the
remnant's westernmost H$\alpha$ filaments and brightest radio emission limb.
Low-dispersion optical spectra of four regions show [\ion{S}{2}]/H$\alpha$
line ratios indicative of shock-heated gas and confirm the absence of
[\ion{O}{3}] emission, suggesting that a relatively low velocity shock ($\leq$70 
km s$^{-1}$) is wide spread throughout most of the remnant.  From reddening
estimates derived from our spectra along with published $\Sigma -D$ distance
estimates, we find G182.4+4.2 likely likely lies at a distance around 4 kpc and thus
is a relatively large (d $\approx50$ pc) and old SNR with an age around 40 kyr
whose faint radio emisison and unusual optical spectra are related to its age, 
low density interstellar environment, and relatively low shock velocity.

\acknowledgements

We thank the referee for many helpful suggestions, Dave Green for comments regarding
G182.4+4.2's faint radio emission, Jessica Klusmeyer for help in obtaining the
optical spectra and the MDM Observatory staff for their excellent instrument
assistance.  This research was made possible by funds from the NASA Space
Grant, the Jonathan Weed Fund, the Denis G. Sullivan Fund, and Dartmouth's
School of Graduate and Advance Studies.

\end{document}